# A better measure of relative prediction accuracy for model selection and model estimation

Chris Tofallis

Reader/Associate Professor of Decision Science

Hertfordshire Business School

c.tofallis@herts.ac.uk

## ABSTRACT

Surveys show that the mean absolute percentage error (MAPE) is the most widely used measure of forecast accuracy in businesses and organizations. It is however, biased: When used to select among competing prediction methods it systematically selects those whose predictions are too low. This is not widely discussed and so is not generally known among practitioners. We explain why this happens.

We investigate an alternative relative accuracy measure which avoids this bias: the log of the accuracy ratio: log (prediction / actual).

Relative accuracy is particularly relevant if the scatter in the data grows as the value of the variable grows (heteroscedasticity). We demonstrate using simulations that for heteroscedastic data (modelled by a *multiplicative* error factor) the proposed metric is far superior to MAPE for model selection.

Another use for accuracy measures is in fitting parameters to prediction models. Minimum MAPE models do not predict a simple statistic and so theoretical analysis is limited. We prove that when the proposed metric is used instead, the resulting least squares regression model predicts the geometric mean. This important property allows its theoretical properties to be understood.

Introduction: What's wrong with MAPE and what causes the problem?

When measuring the accuracy of a prediction the relative the magnitude of relative error (MRE) is often used, it is defined as the absolute value of the ratio of the error to the actual observed value: │(actual – predicted)/actual│ or │(y – ŷ)/y│. When multiplied by 100% this gives the absolute percentage error (APE). This measure is generally only used when the quantity of interest is strictly positive, and we shall make this assumption throughout.
In assessing the accuracy of multiple predictions a way of aggregating is required. If we adopt the arithmetic mean then we shall be using the mean absolute percentage error, MAPE. In some disciplines this is known as the 'mean magnitude of relative error' (MMRE). According to a number of surveys reviewed by Gneiting (2011), MAPE is the most commonly used measure for assessing forecasts in organisations.

According to Kolassa and Martin (2011) "one important problem that has not received adequate attention which arises when MAPE is used as the basis for comparing different methods or systems: using the MAPE for comparisons rewards methods that systematically under-forecast…This problem is poorly understood both among academic forecasters and practitioners in industry". This paper will focus on this problem, explain why it happens, and will propose an alternative accuracy measure. We shall consider its use for model selection as well as its use in fitting parameters to prediction models.

The formula for MAPE is not symmetric in the sense that interchanging ŷ and y does not lead to the same answer, despite the fact that the absolute error is the same before and after the switch. The cause of this asymmetry lies in the denominator of the formula – dividing by the predicted value ŷ instead of the actual y leads to a different result. This issue has been discussed by Armstrong and Collopy (1992), and Makridakis (1993) amongst others, with Makridakis proposing a variation of the MAPE formula to provide symmetry: dividing by the arithmetic mean of the actual and the forecast. This is now known as SMAPE (symmetric mean absolute percentage error) and was originally introduced by Armstrong (1978, p.348). SMAPE has been used as one of the assessment criteria in the M3 forecasting competition (Koehler, 2001). In our simulation experiments we included SMAPE in order to compare its performance with MAPE and two other relative accuracy metrics.

Foss et al. (2003) carried out a detailed simulation study comparing MMRE (MAPE) with other metrics in selecting a forecasting model. Their conclusion was very damning: "The findings suggest that MMRE is an unreliable selection criterion; in many cases, MMRE will select the worst candidate out of two competing models; in particular, MMRE will tend to prefer a model that underestimates to a model that estimates the expected value; in fact, MMRE may be lower (i.e., "better") for a bad model than for a good model even when the good model happens to be the true model"…."MMRE is an unreliable criterion when used to select between competing prediction models… In particular, there is a high probability that MMRE will select a model that provides an estimate below the mean (i.e., "underestimates") to a model that predicts the mean. Given that we prefer information on a precisely defined statistic as the mean to some non-defined, optimistic prediction somewhere between the mean and zero effort, MMRE is inadequate."

Myrtveit and Stensrud (2012) also agree: "MMRE is an *invalid* evaluation criterion because MMRE systematically and consistently ranks models that predict too optimistic [i.e. lower cost] estimates highest".

Miyazaki et al. (1994) used regression based on minimizing MAPE instead of least squares and found that forecasts tended to be too low, a fact which was later confirmed by Lokan (2005). A graph in Foss et al. (2003) shows that there are far more observations above the MAPE regression line than below; we found the same effect, see Figures 1 and 5 later in this paper. Lo and Gao (1997) proposed that the reason for this effect is that for strictly positive data, the range for under-prediction is bounded at zero, whilst that for over-prediction is not. The concept of symmetry for such a situation needs to take this into account. We shall present an approach using the logarithm which makes these two ranges equal in size.

One can in fact better understand the cause of the MAPE bias by noting that points below the line or in the lower part of the data cloud will have a smaller denominator in the percentage error calculation and so will be more influential in pulling the line towards them. By contrast, a regression model based on minimising the mean absolute deviation is used to predict the conditional median, which is why it is sometimes referred to as median regression. Such a line (or curve or surface) will have roughly half the points lying above it and half below it (Appa and Smith, 1973), with a few points lying on the line itself; the median of the errors will be zero. By simple inspection of the formula we see that MAPE regression is a weighted form of median regression, where the weights are $1/y_i$. We now see that points in the lower part of the data cloud will have more weight than the points above them, and so the line will be pulled towards the lower points. Consequently there will be more points above the MAPE line than below. This explains why forecasts based on methods which minimize MAPE will tend to under-predict. Another conclusion is that in any forecasting selection process or competition designed to compare *any* prediction methods, using MAPE as a selection criterion will favour methods which under-predict.

Gneiting (2011) places this stark fact in context: "The absolute percentage error scoring function, which is commonly used by businesses and organizations, and occasionally in academia… tends to support severe under-forecasts…It thus seems prudent that businesses and organizations consider the intended or unintended consequences and reassess its suitability".

Sadly, despite agreement amongst the previously named researchers, most textbooks devoted to business forecasting (e.g. Wilson and Keating (2007), Hanke and Wichern (2009)) make no mention of the above difficulty and so practitioners in business and other organisations are likely to continue using MAPE unless greater there is wider dissemination of its issues.

2. <u>Measuring relative accuracy using the log 'forecast to actual' ratio: Ln Q</u>

Kitchenham et al. (2001) proposed using the ratio of the predicted value to the actual value as a measure of accuracy. We shall denote this by Q, for quotient. Note that Q is the complement of the relative error: 1─ (relative error), and so apart from the shift of one unit, will have the same distribution as the relative error. The complement of the relative error does not seem to have an established name, so Q could be referred to as the 'relative accuracy' or 'accuracy ratio', with 1.0 being the ideal value. Kitchenham et al. (2001) observed that Q was asymmetric because its value is bounded from below by zero, whereas it is unbounded from above. "Since the variable Q is defined on the range 0 to ∞ with a theoretical mean of 1, Q must, by definition, be skewed". We can overcome this asymmetry problem by shall simply take the logarithm. It can then be used as a fitting criterion by applying least squares to Ln Q to produce regression models. It can also be used for comparing the relative accuracy of competing methods by comparing the sum of squares of Ln Q.

Lo and Gao (1997) applied least squares to Ln Q using the Desharnais (1989) data set, comparing the results with those from ordinary least squares and also with least squares applied to the relative error. Their plot of the residuals from Ln(Q) regression showed that about half of the errors were positive and the other half negative. Although a satisfying result, it does not appear to have been influential subsequently, possibly due to lack of theoretical support. We aim to fill this gap by providing some theoretical analysis below. We show that this approach has elegant and useful practical characteristics.

3. Theoretical support for Ln Q

Tornqvist et al. (1985) presented a theoretical study of measures of relative change comparing observations, say *f* and *g* taken at different times or places. They were motivated by asymmetry problems associated with percentages, for example, if B costs 25% more than A, then A is 20% cheaper than B. Similarly, in currency movements: a 25% rise in the value of one currency corresponds to a 20% decrease in another. "These two measures for one single change are often confused, and many mistakes and quarrels could be avoided if only a single measure of relative change were available". Another issue was the lack of additivity of percentages: e.g. an investment that rises by 50% in one period and then falls by 50% in the next, has not returned to its starting price. Likewise, two consecutive rises of 50% do not equate to a single rise of 100%.

The analysis of Tornqvist et al. (1985) can be directly adapted to our context by making *g* represent the predicted value and *f* the observed value. Their work was not concerned with measurement error, accuracy, forecasting or goodness of fit, and this paper is the first instance where these valuable results have been brought into the field of forecasting.

They considered ten measures, which included:

$(f-g)/f$, $\quad$ $(f-g)/g$, $\quad$ $(f-g)/[\tfrac{1}{2}(f+g)]$, $\quad$ Ln $(g/f)$

$(f-g)/(fg)^{1/2}$, $\quad$ $(f-g)/\min(f,g)$, $\quad$ $(f-g)/\max(f,g)$

We immediately see that these correspond to some measures of prediction accuracy as proposed for use in forecasting. The first of these measures is the relative error, the second is MER - the error relative to the predicted value - it has the opposite problem to MRE in that it tends to over-predict when used for fitting (Lokan, 2005). The third corresponds to SMAPE. The last two measures have appeared in the cost estimation literature under the names of balanced relative error, (Miyazaki et al. 1991), and inverted balanced relative error (Miyazaki et al. 1994). The former compares the error relative to the actual value in the case of over-estimation, and relative to the predicted value for under-estimation. The inverted balanced relative error does the reverse of this.

Tornqvist et al. (1985) considered the analytic properties of their ten measures, and looked for symmetry, additivity, continuity, and homogeneity. Their valuable result was a clear conclusion: "If we require, as is natural in economics, that the indicators of relative change be symmetric, additive, and normed, we are left with the indicator $\ln(g/f)$". They refer to this measure as the log difference or log change, because $\ln(g/f) = \ln(g) − \ln(f)$.

This measure is symmetric in the sense that interchanging *f* and *g* merely alters the sign:

ln(predicted/actual) = ln(predicted) – ln(actual)

= – [ln(actual) – ln(predicted)] = –ln(actual /predicted)

Remarkably, it provides within the same metric both a ratio and a difference!

We can compare it with percentage error using a simple example from Shepperd and MacDonell (2011): one project is predicted to cost 10 but actually costs 100; conversely, another project is predicted to cost 100 but ends up costing 10. For these projects the magnitudes of the percentage errors are 90% and 900% respectively - a huge difference. By contrast, using the proposed metric gives the same magnitude of accuracy for each case.

### 3.1 Least squares models based on Ln Q predict the geometric mean

Consider having multiple (positive) observations of y for a given x. Consider representing these by a single value, or measure of location Y, such that the sum of squares of $\ln(Q_i)$ is minimized:

Min $\Sigma$ [ ln (Y/$y_i$) ]$^2$   or   Min $\Sigma$ [ ln (Y) – ln($y_i$) ]$^2$

Differentiating with respect to Y and setting to zero:

2 $\Sigma$ [ ln (Y/$y_i$) ]/Y = 0

so $\Sigma [ \ln (Y/y_i) ] = 0$

and $\Sigma [ \ln (Y) - \ln(y_i) ] = 0$

thus $n \ln(Y) = \Sigma \ln (y_i)$

and $Y = \exp[1/n \ \Sigma \ln (y_i) ] = [\Pi (y_i) ]^{1/n}$ = geometric mean

Thus our estimator Y is the geometric mean. This derivation can be generalised to the case where Y is replaced by a function of x, we then have a means of fitting models to data which predict the geometric mean. This will be a superior alternative method for our purposes because estimating parameters by minimising MAPE leads to models which under-predict and where the prediction is not a simple statistic or measure of location.

The arithmetic mean of a set of numbers is always greater than the geometric mean. OLS regression models predict the arithmetic mean but are known to be affected by outliers. Least squares applied to log $Q_i$ will predict a value below the arithmetic mean and will be less affected by outliers – a second useful property.

We can extend the argument to continuous distributions in y (y>0). The Appendix shows that regression models based on applying least squares to log Q would be expected to predict the geometric mean for any continuous distribution (y>0) conditioned on x.

### 3.2 Unbiasedness of predictions

A prediction system would be considered biased if its forecasts are systematically too high or too low. The point of reference is usually the arithmetic mean. Thus OLS (ordinary least squares) regression is said to be mean-unbiased because it predicts the arithmetic mean of y for a given x. However, we know the arithmetic mean is sensitive to unusually large observations, so this type of unbiasedness may not be ideal. If the median estimator is preferred then least absolute error regression is appropriate because it is median-unbiased. Given that our proposed method predicts the geometric mean, we can say that it is geometric mean unbiased.

We have seen that the regression method we are discussing predicts the geometric mean of y for a given x-value: $Y_{pred}$ = geomean{ $y_i$ }

Now, taking the $n^{th}$ power: $(Y_{pred})^n = \Pi\, y_i$

And dividing by the right hand side: $\Pi\, (Y_{pred}/y_i) = 1$

Thus the product (and geometric mean) of the accuracy ratios is unity. So the predictions will be unbiased in relative terms in this sense. This implies, for example, that a prediction with a Q ratio of 5/4 is counter-balanced by one with a ratio of 4/5; i.e. a prediction which is 25% too high is balanced by one which needs to be increased by 25% (to get from 4/5 to 1). This can be viewed as a form of symmetry which is appropriate for relative metrics.

Next consider the predictions for all observed x-values, we can prove that our regression will result in the geometric mean of all accuracy ratios being equal to unity, provided that the model equation includes a constant coefficient, i.e. $\hat{y} = a\, f(x;b)$.

This is analogous to the OLS property that the arithmetic mean of the residuals is zero provided the model includes a constant term.

Proof: To minimize $\Sigma\, [\, \ln(\,a\, f(x_i;b)\, /\, y_i\,)\, ]^2$ or $\Sigma\, [\, \ln(a\, f(x_i;b)) - \ln y_i\,)\, ]^2$

requires that the derivative with the respect to *a* is zero:

$2\, \Sigma\, [\, \ln(\,a\, f(x_i;b)\,) - \ln y_i\,)\, ]/[\,a\, f(x_i;b)] \cdot \partial\, a f(x_i;b)\, /\partial a = 0$

Therefore $1/a\, \Sigma\, [\, \ln(\,a\, f(x_i;b)\,) - \ln y_i\,)\, ] = 0$

and since *a* cannot equal zero, we have: $\Sigma\, \ln[\,a\, f(x_i;b)\, /\, y_i\,] = 0$

Taking antilogs gives: $\Pi\, (\hat{y}/y_i) = 1$ or $\Pi\, Q_i = 1$ and so geometric mean =1   QED

Multiplying together all ratios of predicted/actual is a very simple way of aggregating individual accuracy measurements which takes into account the relative size of the observations and which gives an easily understandable result: unity.

If we take the logs: $\Sigma\, \ln Q_i = 0$ or $\Sigma\, [\ln \hat{y} - \ln y_i\,] = 0$

Thus the arithmetic mean of the log of the accuracy ratio equals zero.

### 3.3 Connection with multiplicative error regression

In this section we demonstrate an important connection between Ln Q and multiplicative error models. Consider the general class of models with a multiplicative error factor ε:

y = ŷ ε, where the prediction ŷ is a function of x and some parameters.

Consider applying least squares to Ln $Q_i$ :

$\Sigma$ [ln Q]$^2$ = $\Sigma$ [ln(ŷ / ln y)]$^2$ or $\Sigma$ [ln (ŷ) – ln y]$^2$ = $\Sigma$ [ln y– ln (ŷ)]$^2$ = $\Sigma$ [ln ε]$^2$

Thus our method is identical to applying least squares to the log of the multiplicative error.

In the previous section we showed that Π (ŷ /$y_i$ ) = 1 which implies Π ($y_i$/ ŷ) = 1.

Hence Π $\varepsilon_i$ = 1, the geometric mean of the error factors will be unity for the proposed form of regression.

Multiplicative error models are well-suited to heteroscedastic data.

### 4. Model fitting: Empirical applications

We now apply the estimation method to a data set collected in Finland by Hannu Maki for the TIEKE organization (downloadable from promisedata.org). It consists of 38 observations of project effort (y) measured in person-hours together with the sizes of the projects (x) measured in function points (FP). The project efforts range from 460 hours to 23,000 hours and it is apparent from Figure 1 that as project size grows so does the scatter or spread of project effort, i.e. there is clear evidence of heteroscedasticity, which is the norm for this type of data as reported by the MERMAID project and many others. According to Stensrud et al, 2003, MAPE "is the *de facto* standard evaluation criterion to assess the accuracy of software project prediction models". We shall compare results from fitting least squares models based on Ln Q with models based on minimizing MAPE. This is also appropriate because both methods focus on relative accuracy whereas OLS does not.

We begin with a simple straight line relationship between effort and FP. Minimizing MAPE led to the following model: Predicted effort = 10.05 + 3.8 FP

Whereas with our approach we found: Predicted effort = 52.93 + 7.525 FP

These lines are displayed in Figure 1 and it is apparent that the MAPE approach does not pass through the cloud of points very well – its position is too low. This illustrates the bias of MAPE.

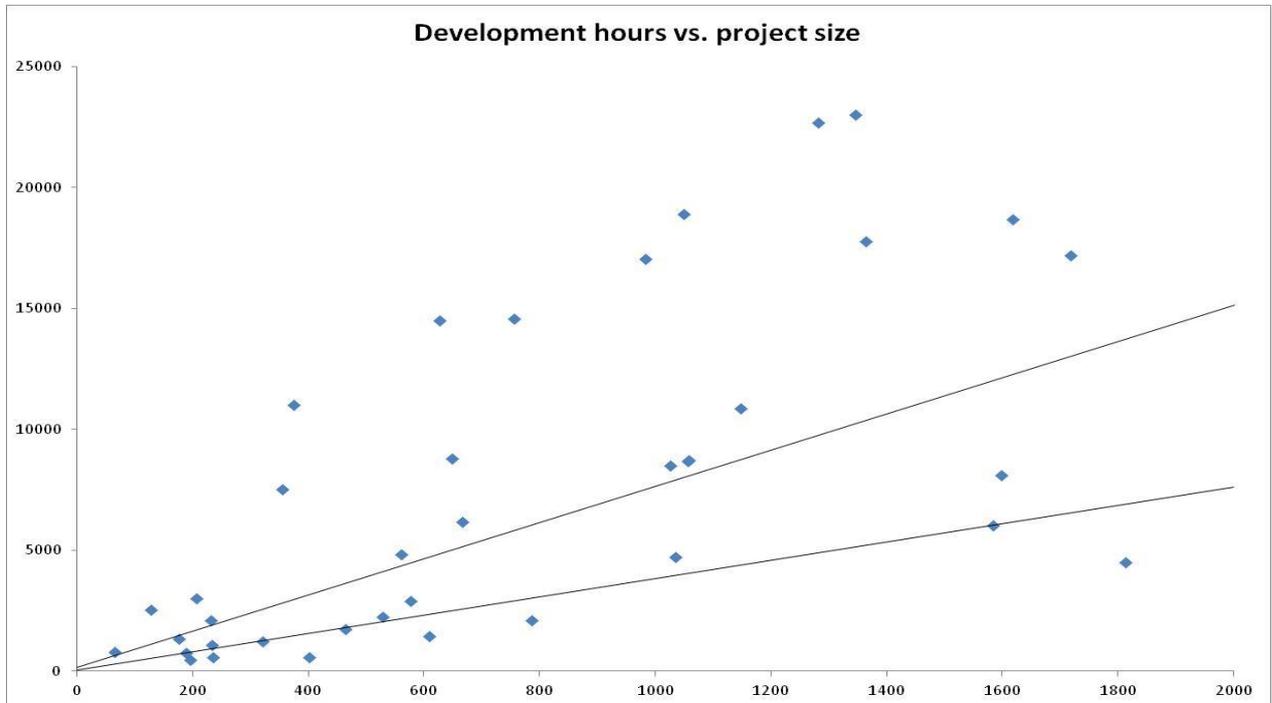

Figure 1. Plot of project development time versus size of project. The lower line has been derived by minimizing the MAPE, the upper line using the proposed method.

Another way of comparing results is to use the accuracy ratios (predicted effort/actual effort), ideally equal to unity, and so the log accuracy ratio is ideally zero. Plotting this quantity against each project's size (FP) enables us to compare the performance of the models. On the bar charts a positive bar implies an over-prediction.

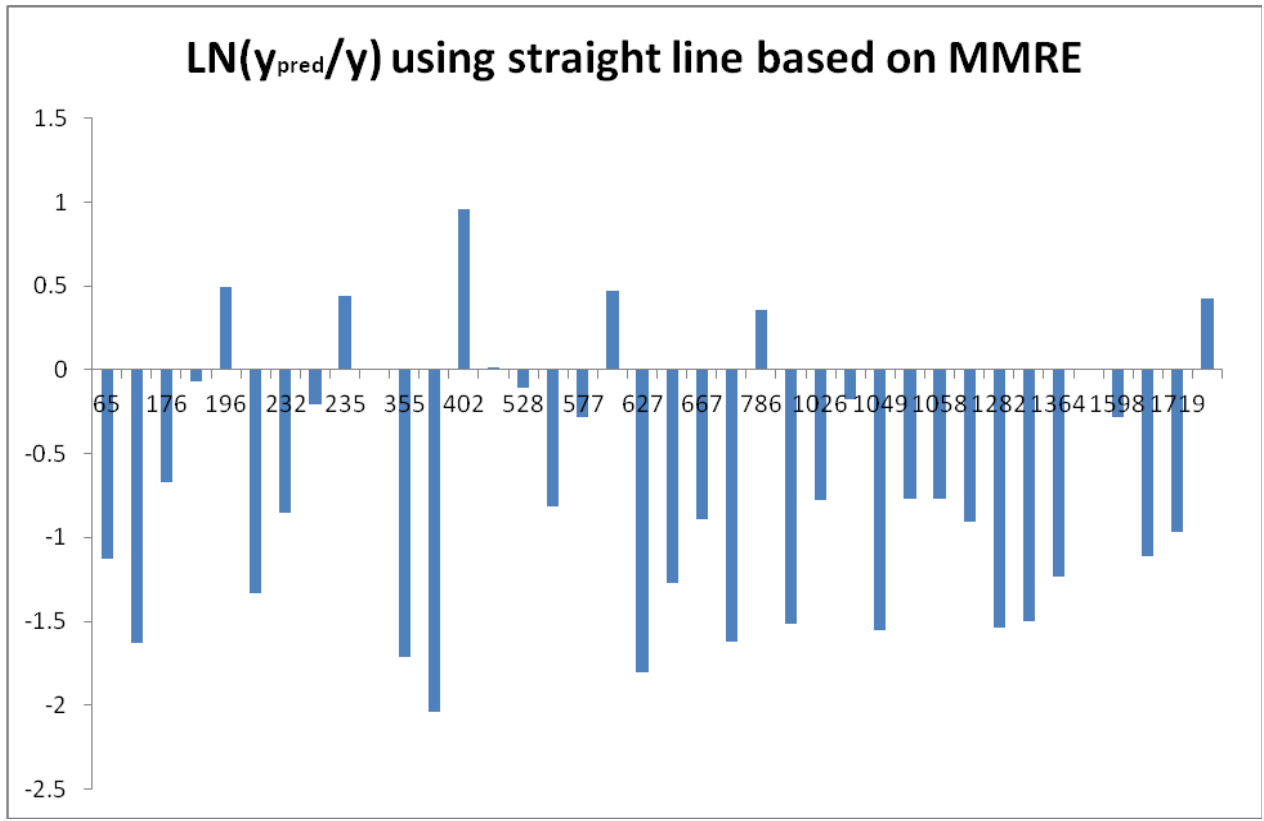

Figure 2. Log of accuracy ratio versus size of project: straight line model based on minimizing MAPE.

We see that the MAPE-based results (Figure 2) mostly consist of under-predictions, with very few over-predictions. By contrast when the proposed method is used there is greater symmetry (Figure 3). We also confirmed that the mean of the plotted statistic was equal to zero (equivalent to the geometric mean of the accuracy ratios equating to unity) – not only is this a valuable property in itself but it also acts as a useful check on the computations.

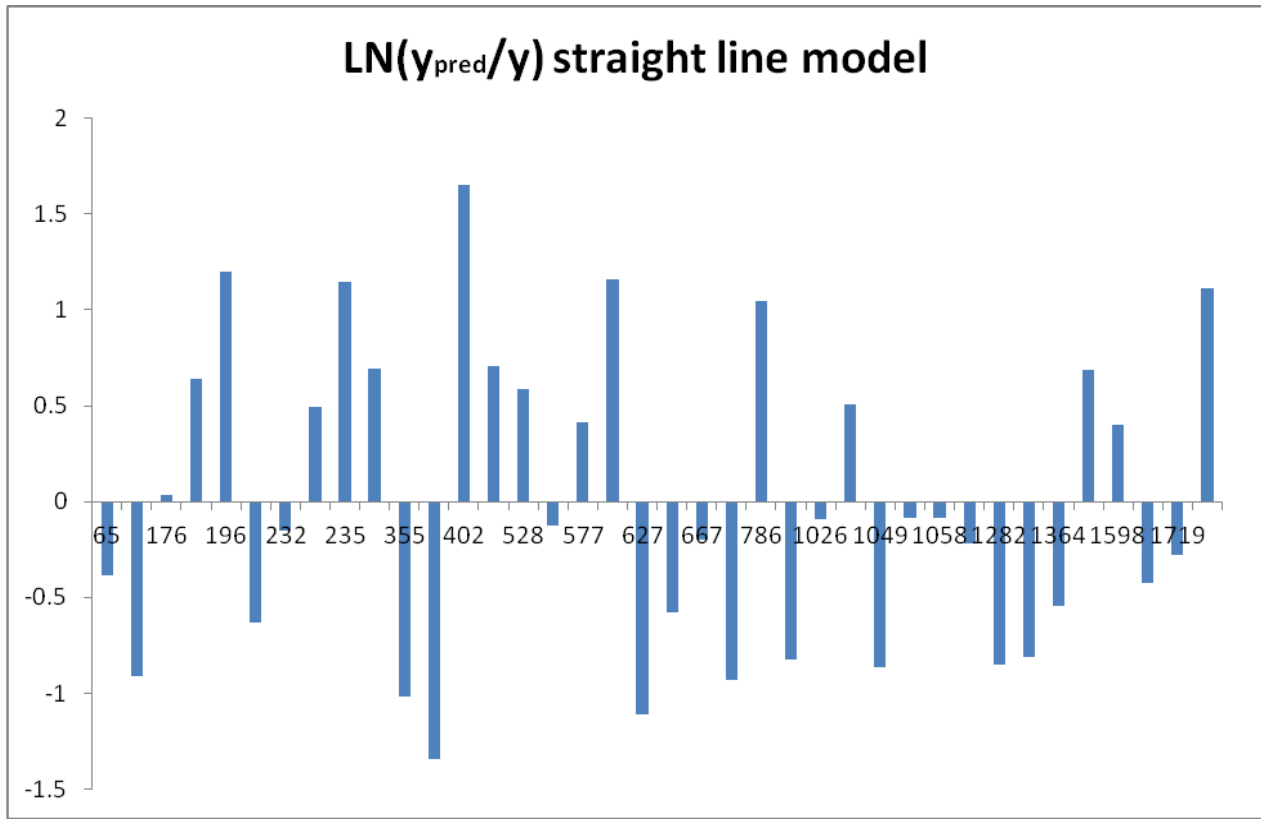

Figure 3. Log of accuracy ratio versus size of project: straight line model with least squares applied to Ln $Q_i$.

We then fitted a curved relationship: a power function. The best-fit MAPE model was:

Predicted effort = 0.892 (FP)$^{1.235}$

whereas when it was fitted using least squares applied to Ln Q we obtained:

Predicted effort = 1.70 (FP)$^{1.053}$

Figures 4 and 5 display the log accuracy ratios. It is again apparent that when we minimize the MAPE the results are not symmetric: they are mostly under-predictions, with very few over-predictions. Once again the proposed method displays more symmetry (Figure 5). It is the use of logs that overcomes the asymmetry issue that Kitchenham et al. (2001) observed in the Q accuracy measure.

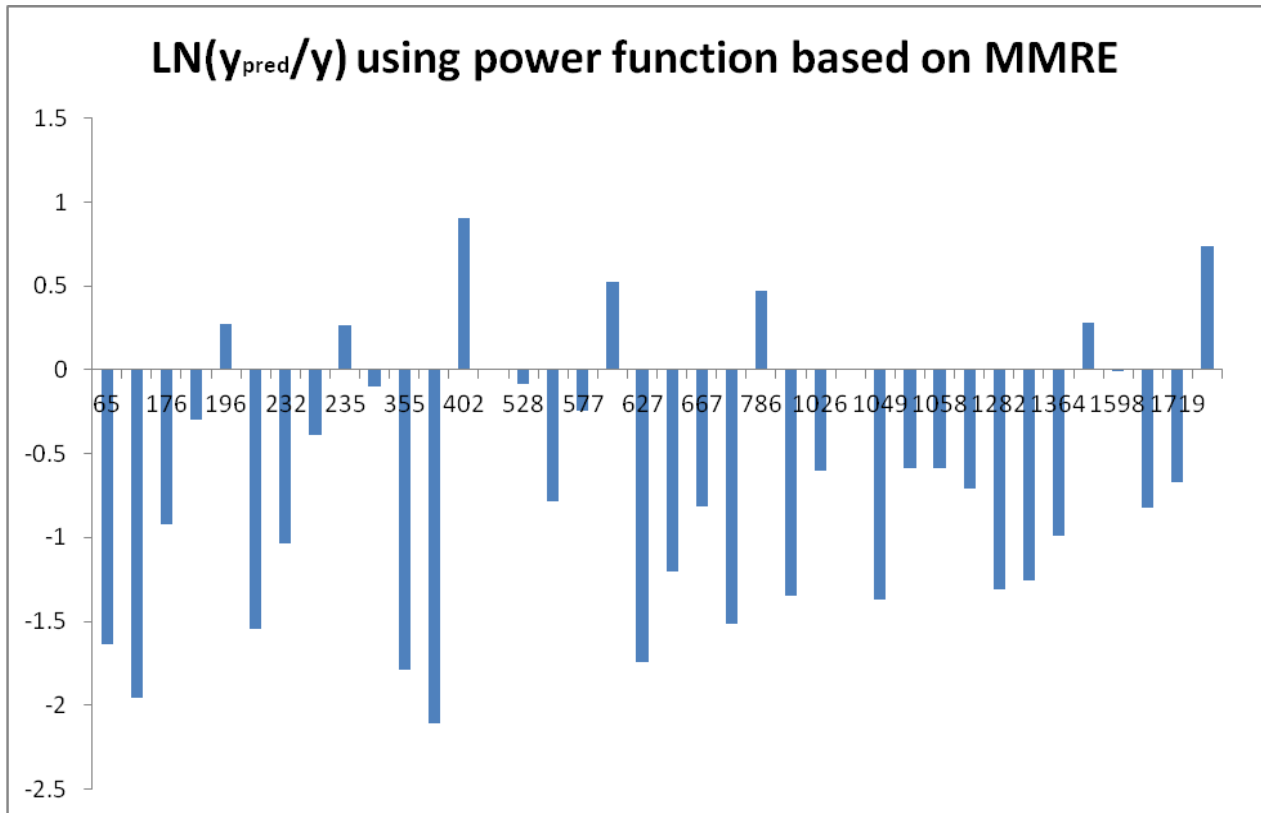

Figure 4. Plot of log accuracy ratio versus project size using a power function prediction model fitted by minimizing MAPE.

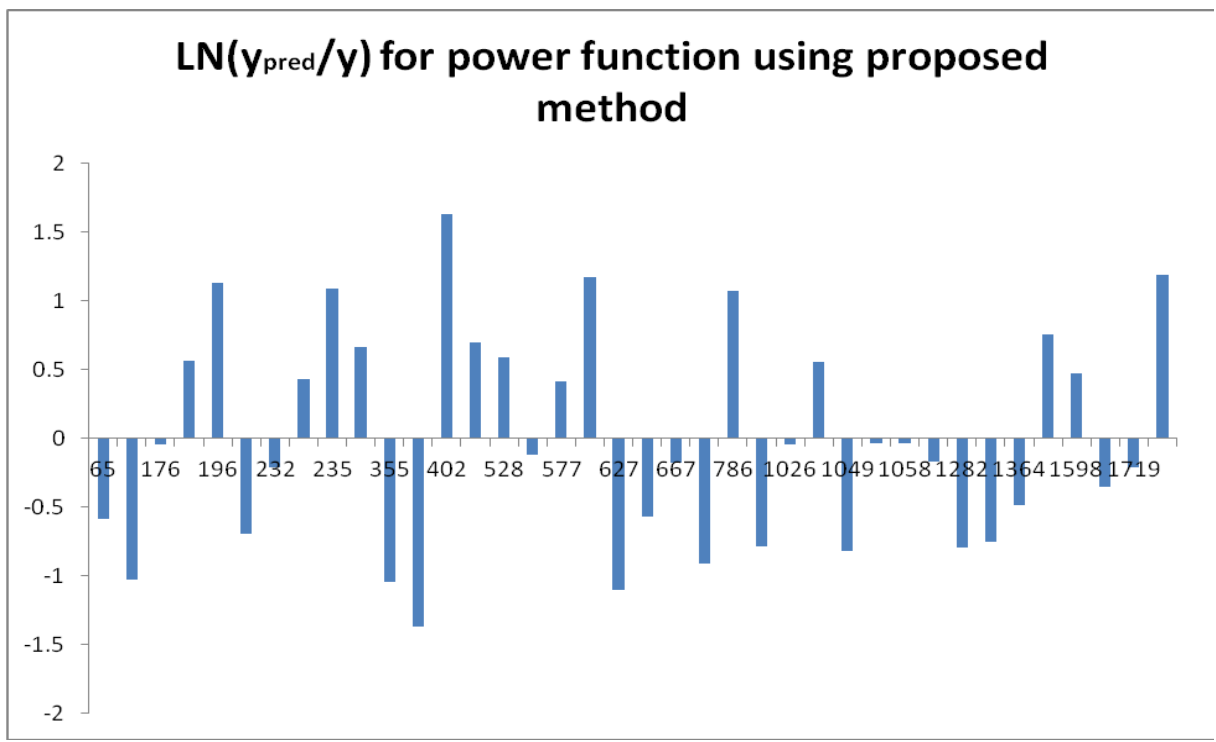

Figure 5. Plot of log accuracy ratio versus project size using a power function model fitted by least squares applied to Ln $Q_i$.

Second application: The Desharnais dataset consists of 81 observations of software projects ranging from 546 hours to 19,894 hours; the function point range is 62 to 1,116. It has been studied by various researchers (e.g. Foss et al, 2003, Kitchenham et al, 2001) and is freely downloadable from the PROMISE.org data repository. The investigation by Lo and Gao (1997) looked at three methods for fitting a model to the data: OLS, least squares based on percentage errors (described in detail in Tofallis, 2008), and the Ln Q approach which they proposed. They demonstrated the superiority of the latter (using a power law model) in terms of neither over-predicting nor under-predicting. Since they did not fit a model which minimized the MAPE, we shall do that here. For a straight line model the lines are displayed in Figure 6. As expected the line based on MAPE is positioned too low, in fact it under-predicts 61 of the 81 projects, whereas the other line under-predicts 49 projects. Consequently, and not surprisingly, plots of the log accuracy ratio possess the same general features as seen above for the Finnish data set i.e. randomly scattered around zero with no evidence of asymmetry or trend. (Those plots are not displayed here for reasons of space.)

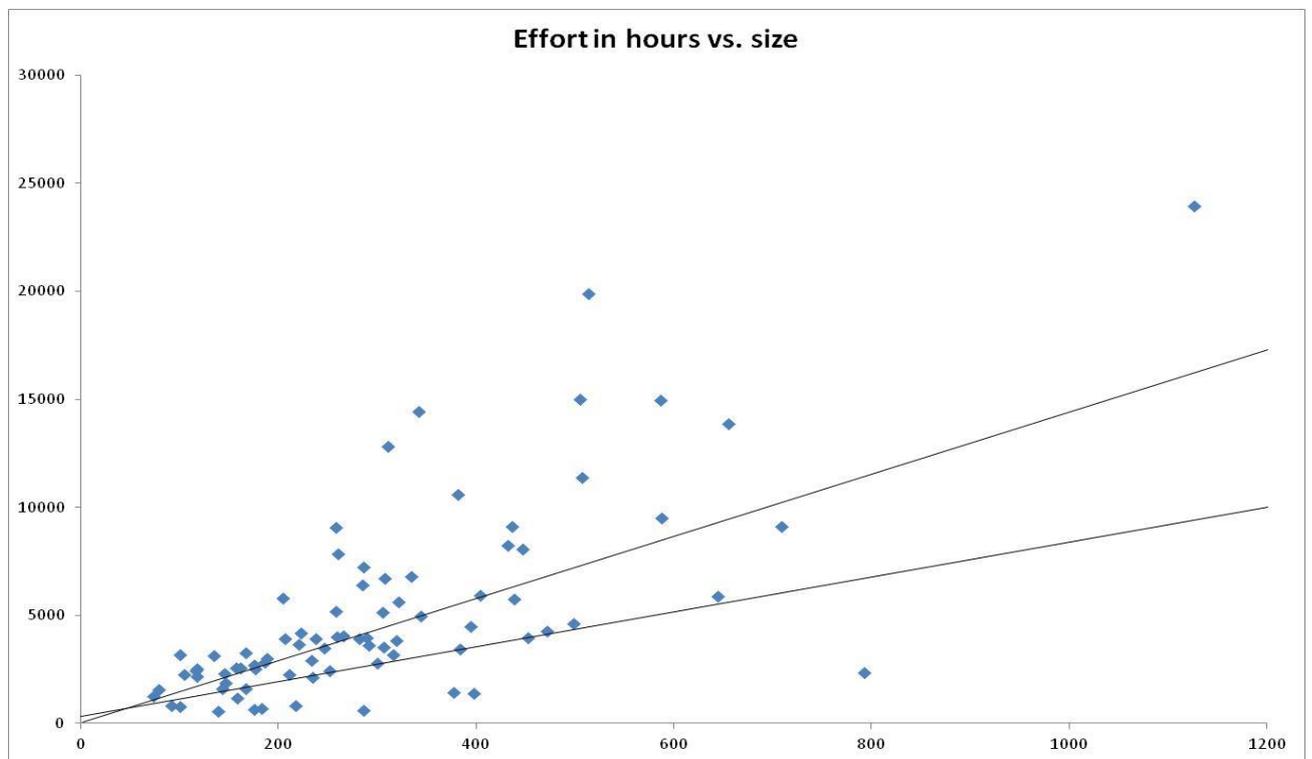

Figure 6. Desharnais data set. The upper line was computed using the proposed method and the lower line is that which minimizes MAPE.

## 5. Model selection: Simulation experiments

In order to compare the performance of different measures of relative accuracy we carried out Monte Carlo simulations to see which measure was best at identifying the model that generated the data from five competing models. In this way we tested the model selection capability of each metric. This is a suitable approach given that accuracy measures are widely used for selecting among forecasting systems. We followed the approach of Foss et al (2003) who generated data using the model:

$$y = e^\alpha x^\beta \varepsilon$$

with $\varepsilon$ log-normally distributed with mean unity. They chose this model because it was representative of datasets on software project duration (such as that of Desharnais, 1989) in terms of heteroscedasticity and nonlinearity. Note the use of a multiplicative error *factor* to model heteroscedastic variation. In each case thirty data points were generated with x ranging from 50 to 1500 in steps of 50, and y as above with $\alpha = 3.03$ and $\beta = 0.943$. Two of their competing models were given parameters so as to under-predict the true model, and the other two models were designed to give higher predictions. They carried out separate pairwise comparisons between the true model and one alternative model, whereas we simultaneously compared the true model with four alternatives. In this way we found that three of the alternative models used by Foss et al (2003) were rarely picked out as best by any criterion and so were effectively redundant. We therefore chose the parameters of our alternative models so that they would provide more 'competition': $\alpha$ all the same, $\beta = 0.92, 0.93, 0.95,$ and 0.96.

The five competing models were as follows:

TRUE MODEL: $y = e^{3.03} x^{0.943}$

MODEL 1: $y = e^{3.03} x^{0.92}$

MODEL 2: $y = e^{3.03} x^{0.93}$

MODEL 3: $y = e^{3.03} x^{0.95}$

MODEL 4: $y = e^{3.03} x^{0.96}$

One other small departure from Foss et al (2003) was that their error factor had an *arithmetic* mean of unity, and in order to achieve this had to tie this value to the variance of the lognormal distribution. We instead used a *geometric* mean of unity which is more natural for a multiplicative error and this allowed the variance to be set independently of the mean. We also increased the number of samples of thirty data points from one thousand to ten thousand and in each case recorded the number of times that the true model was chosen by each selection criterion as being closest to the data. We also experimented with the amount of noise in the data by varying the standard deviation of the error. The results are displayed in Table 1.

| Noise level ($\sigma$) | Percent correct using MAPE | Percent correct using $\Sigma (\ln Q)^2$ | Percent correct using LSD | Percent correct using SMAPE |
|---|---|---|---|---|
| 0.1 | 86% | 88% | 82% | 82% |
| 0.2 | 43% | 59% | 48% | 52% |
| 0.3 | 19% | 43% | 28% | 35% |
| 0.4 | 7% | 34% | 16% | 27% |

TABLE 1. Performance of different accuracy measures in identifying the true underlying power function model which generated the data, at different noise levels.

As expected, when the noise in the data is small the performance is good. However, as the noise level increases MAPE is clearly seen to be the worst performer.

The third column of the table gives the performance of another metric which was found to perform very well by Foss et al. (2003) in that it correctly chose the correct model more often than the other criteria they studied. They refer to this as LSD (logarithmic standard deviation) and it has some similarity with the measure we have been discussing:

$$\text{LSD} = \sqrt{\frac{\Sigma (0.5\, s^2 - \ln Q_i)^2}{n-1}}$$    where $s^2$ is the variance of the $Q_i$.

As can be seen from the table, LSD was a clear improvement on MAPE but did not perform as well as the simpler $\Sigma (\ln Q_i)^2$ formula.

We then proceeded to carry out a further set of simulations, this time we simplified matters by removing any dependence on x and investigated how well the three measures performed when the underlying signal was a constant with noise:

$y_i = c \, \varepsilon_i$ with $\varepsilon_i$ log-normally distributed with geometric mean equal to unity, this is achieved by setting the mean of the underlying normal distribution to zero, whilst the standard deviation is varied to adjust the noise level. We set c=10, and the competing models were y = 8, 9, 10, 11, and 12 respectively.

The results are displayed in Table 2.

| Noise level ($\sigma$) | Percent correct using MAPE | Percent correct using $\Sigma (\ln Q)^2$ | Percent correct using LSD | Percent correct using SMAPE |
| --- | --- | --- | --- | --- |
| 0.1 | 97% | 100% | 98% | 98% |
| 0.2 | 57% | 81% | 72% | 75% |
| 0.3 | 27% | 62% | 45% | 54% |
| 0.4 | 11% | 52% | 29% | 39% |

Table 2. Performance of different accuracy measures in identifying the true underlying model y =10 which generated the data, with different *multiplicative* noise levels.

Table shows that $\Sigma (\ln Q)^2$ performs best out of the four metrics, moreover this is the case at all noise levels. SMAPE is second best, and LSD third. Once again MAPE is clearly the worst performer. Table 3 shows the proportion of the time that each metric chooses a model that predicts too low or too high respectively. As expected, MAPE favours models which predict too low. By contrast, when LSD does not choose the correct model it has the opposite bias in that it selects models which over-predict; though this bias is not as strong as that of MAPE. As regards bias, SMAPE and $\Sigma (\ln Q)^2$ seem to be fairly neutral.

| Noise level (σ) | MAPE | | Σ (ln Q)² | | LSD | | SMAPE | |
|---|---|---|---|---|---|---|---|---|
| | Under-estimate model chosen | Over-estimate model chosen | Under-estimate model chosen | Over-estimate model chosen | Under-estimate model chosen | Over-estimate model chosen | Under-estimate model chosen | Over-estimate model chosen |
| 0.1 | 3% | 0% | 0% | 0% | 0% | 2% | 2% | 0% |
| 0.2 | 41% | 2% | 9% | 10% | 3% | 25% | 11% | 14% |
| 0.3 | 69% | 4% | 18% | 20% | 4% | 51% | 21% | 25% |
| 0.4 | 88% | 1% | 23% | 25% | 4% | 67% | 31% | 30% |

Table 3. Bias of different accuracy measures as indicated by their tendency to select models that over-predict or under-predict. Same simulation setup as for Table 2.

Next, we carried out simulations where the generated data model had an *additive* error term to represent the noise:

$y_i = c + e_i$   with $e_i$ normally distributed with mean zero and standard deviation σ. As before we set c = 10 with the same competing models y = 8, 9, 10, 11, and 12 respectively.

The results in Table 4 show that LSD is best at identifying the true model when the error is additive, Σ (ln Q)² and SMAPE are almost as good at low noise but gradually worsen, though they are clearly always much better than MAPE.

| Noise level (σ) | Percent correct using MAPE | Percent correct using Σ (ln Q)² | Percent correct using LSD | Percent correct using SMAPE |
|---|---|---|---|---|
| 1.0 | 97% | 100% | 100% | 98% |
| 1.5 | 78% | 90% | 92% | 87% |
| 2.0 | 54% | 76% | 82% | 74% |
| 2.5 | 34% | 60% | 72% | 64% |

Table 4. Performance of different accuracy measures in identifying the true underlying model y =10 which generated the data, with different *additive* noise levels.

| Noise level ($\sigma$) | MAPE | | $\Sigma (\ln Q)^2$ | | LSD | | SMAPE | |
|---|---|---|---|---|---|---|---|---|
| | Under-estimate model chosen | Over-estimate model chosen | Under-estimate model chosen | Over-estimate model chosen | Under-estimate model chosen | Over-estimate model chosen | Under-estimate model chosen | Over-estimate model chosen |
| 1.0 | 3% | 0% | 0% | 0% | 0% | 0% | 0% | 2% |
| 1.5 | 20% | 2% | 8% | 2% | 3% | 5% | 6% | 7% |
| 2.0 | 42% | 0% | 20% | 4% | 7% | 11% | 12% | 14% |
| 2.5 | 54% | 1% | 34% | 15% | 11% | 17% | 16% | 20% |

Table 5. Bias of different accuracy measures as indicated by their tendency to select models that over-predict or under-predict. Same simulation setup as for Table 4.

It is worth noting that $\Sigma (\ln Q)^2$ has a bias toward under-prediction when the noise is additive, and we can explain this using two facts. Firstly, this measure is geared to choosing whichever model predicts closest to the geometric mean of the data, and secondly, if a set of identical numbers (e.g. $y_i = 10$) is subjected to a mean-preserving spread — i.e. are spread apart from each other while leaving the arithmetic mean unchanged (e.g. by adding $N(0, \sigma)$ errors) — then the geometric mean always decreases (this is a consequence of the arithmetic mean – geometric mean inequality).

Note that for any prediction method $\hat{y} = f(x)$ the conditional value for a given x is a constant value, and so the above analysis can be generalised. If the data is being generated by a model of the form $y_i = f(x) + e_i$ where the error term follows *any* symmetric distribution with mean zero, then we would ideally like to discover the prediction system $\hat{y} = f(x)$, but notice that this implicitly assumes an *arithmetic* mean being applied to $f(x) + e_i$ (imagine we had multiple y-values for a given x). By contrast using a geometric mean on multiple values of $f(x) + e_i$ for a given x will lead to a lower value than the arithmetic mean. (The ideal accuracy metric for selecting the arithmetic mean is the mean square error.)

If instead the true underlying model involves a multiplicative error (and is therefore heteroscedastic): $\hat{y} = f(x) \varepsilon_i$ then ideally we would like to discover the prediction system $\hat{y}$

= f(x) which this time implies a geometric mean is being assumed when analysing multiple y-values for a given x.

Since $\ln(y_i) = \ln(f(x)) + \ln(\varepsilon_i)$ the previous argument leads to $\ln(\varepsilon_i)$ being distributed according to any symmetric distribution with arithmetic mean zero, so that the geometric mean of the $\varepsilon_i$ is zero. In such cases the best metric to use for model selection is $\Sigma (\ln Q)^2$.

In summary, based on our findings we would make the following recommendations:

- If the data appears to have a symmetric additive error i.e. is homoscedastic, and relative errors are of more concern than absolute errors, then use LSD.
- If the data appears to have a multiplicative error i.e. is heteroscedastic, then use $\Sigma (\ln Q)^2$ to select your prediction method.

6. <u>Summary and conclusion</u>

Given the diversity of methods available, 'Which forecasting method is best?' is a question that is bound to arise. Myrtveit et al. (2005) were concerned about the validity and reliability of studies which aimed to answer this question. Their conclusion was: "Empirical studies … do not converge with respect to the question. The reason for this lack of convergence is poorly understood." They and others have been perplexed by the fact that one accuracy indicator may select one prediction system, whilst another accuracy indicator will predict another:

> "Our study thus suggests that the conclusions on "which model is best" to a large extent will depend on the accuracy indicator chosen. This is a serious problem because, at present, we have no theoretical foundation to prefer, say, MMRE to MMER [magnitude of error divided by estimate] or MAR [mean absolute residual] to MBRE [magnitude of balanced relative error]."

Kitchenham et al. (2001) noted that: "in a given situation different accuracy statistics often give contradictory results. This indicates that they are not measuring the same aspect of prediction accuracy. We believe that the lack of understanding of what different accuracy statistics actually measure is hindering progress".

From these quotations there is clearly some confusion regarding why different accuracy measures do not lead to the same prediction method being selected. One way forward is to recognise that the predicted quantity has randomness associated with it, and so under slightly different circumstances different results could be observed for a given x-value. If we actually had a multiplicity of such results we would be faced with a choice of what is the typical value i.e. which measure of location to use. Each accuracy measure gives access to a different measure of location. The following table presents three options, each of which leads to a different measure of location or expected value. Thus the confusion is really related to the fact that there is a variety of location measures.

| Accuracy measure or Loss function | Accuracy type | Regression model | Expected value | Unbiasedness property |
|---|---|---|---|---|
| $\Sigma \|e_i\|$ | Absolute | Additive error | Median | Median error = 0 |
| $\Sigma e_i^2$ | Absolute | Additive error | Arithmetic mean | Mean error = 0 |
| $\Sigma (\ln Q)^2$ | Relative | Multiplicative error | Geometric mean | $\Pi (\hat{y}_i / y_i) = 1$ |

Table 6. Properties of three accuracy measures.

It is easily forgotten that we do not have to use just one approach – just as we are not restricted to reporting a single measure of location. Each prediction method gives us a different piece of information, and the more information we have the better off we are. Of the three measures in the table, only the last is based on *relative* error. It also has the property of not being as affected by large outliers as OLS. (The famous arithmetic mean-geometric mean inequality guarantees that the geometric mean cannot exceed the arithmetic mean.) Hence, whilst using the MAPE criterion in regression will tend to under-predict, and OLS may over-predict, the proposed approach offers an intermediate figure. It also produces forecasts which have a form of unbiasedness which is appropriate for relative accuracy: the product of the forecast to actual ratios equates to unity.

The approach has an important connection with multiplicative error regression, which is natural given that we are dealing with a relative measure of accuracy.

Ln(Q) = ln(forecast/actual) is dimensionless and can therefore be used in comparisons across data sets in the way that average percentage errors are often informally used. Another benefit of Ln(Q) over percentage error, is that it treats positive and negative errors with an appropriate symmetry, and the possible ranges are equal on each side. Specifically, if we switch the actual and predicted values in the formula then this merely reverses the sign. (We recommend plots of $Ln(Q_i)$ in assessing models where relative accuracy is important.) Green and Tashman (2009) surveyed forecasters on which denominator they preferred in their relative error, with 56% preferring the actual value as denominator. Using the logarithm brings the benefit of removing this issue. Also, the proposed approach estimates a known statistic – the geometric mean – whereas minimizing MAPE or SMAPE do not produce an established measure of location.

To summarise: This paper explained why MAPE regression leads to predictions that are too low and considered an alternative based on the log of the accuracy ratio, Ln(Q). A new connection was made between forecasting and Tornqvist et al's (1985) theoretical work on measuring change which showed that Ln(Q) is the only form known to satisfy the set of desirable properties which they specified. We also proved that applying least squares regression to this measure led to predictions of the geometric mean, and also that it is equivalent to multiplicative error regression. Aside from these theoretical reasons, we also provided evidence from simulations to demonstrate the superiority of $\Sigma (\ln Q)^2$ over MAPE as a measure for selecting a prediction system. The simulations also showed that this measure is better than two other proposed alternative metrics: SMAPE and LSD, for model selection when data is heteroscedastic.

In conclusion, using MAPE as a fitting criterion results in a majority of predictions being too low, whereas applying least squares to Ln Q does not suffer from this bias. Even if MAPE is not used for model estimation, but only at the model selection stage, it will still be problematic because it will favour models whose forecasts are too low. We therefore conclude that for strictly positive data, the proposed metric appears to be superior to MAPE as a measure of relative prediction accuracy.

# APPENDIX

We demonstrate that least squares applied to log Q gives the geometric mean as the expected value for any continuous distribution (y>0). This is analogous to the result that the value which minimises the mean squared error over a distribution is the arithmetic mean. These results can be viewed as connections between loss functions and measures of location.

To begin, note that for a continuous distribution f(y), the geometric mean G is defined by

$\log G = \int f(y) \log y \, dy$.

f(y) could be the conditional distribution of y for a given x-value.

The associated regression metric for our log accuracy ratio is:

$\int f(y) [\log Y - \log y]^2 \, dy$

Differentiating with respect to Y:

$2 \int f(y) [\log Y - \log y] / Y \, dy$

Setting this to zero:

$\log Y \int f(y) \, dy = \int f(y) \log y \, dy$

Since $\int f(y) \, dy = 1$ we have $\log Y = \int f(y) \log y \, dy$

Which implies Y is the geometric mean as required.